# Quantum Fourier transform spectroscopy of biexciton

Hiroya Seki[1], Kensuke Miyajima[2], and Ryosuke Shimizu[1*]

[1]Department of Engineering Science, Graduate School of Informatics and Engineering, The University of Electro-Communications, 1-5-1 Chofugaoka, Chofu, Tokyo 182-8585, Japan
[2]Department of Applied Physics, Graduate School of Science, Tokyo University of Science, 6-3-1 Niijuku, Katsushika-ku, Tokyo 125-8585, Japan

**Abstract**

Fourier transform spectroscopy with classical interferometry corresponds to the measurement of a single-photon intensity spectrum from the viewpoint of the particle nature of light. In contrast, the Fourier transform of two-photon quantum interference patterns provides the intensity spectrum of the two photons as a function of the sum or difference frequency of the constituent photons. This unique feature of quantum interferometric spectroscopy offers a different type of spectral information from the classical measurement and may prove useful for nonlinear spectroscopy with two-photon emission. Here, we report the first experimental demonstration of two-photon quantum interference of photon pairs emitted via biexcitons in the semiconductor CuCl. Besides applying Fourier transform to quantum interference patterns, we reconstruct the intensity spectrum of the biexciton luminescence in the two-photon sum or difference frequency. We discuss the connection between the reconstructed spectra and exciton states in CuCl as well as the capability of quantum interferometry in solid-state spectroscopy.

**Main**

Interference of the quantum field of multiple photons involves versatile phenomena in quantum science and technology. The first experimental demonstration, known as the Hong-Ou-Mandel (HOM) effect, was reported in 1987 [1]. Since then, this phenomenon has been widely used, ranging from the testing of the fundamental concepts of quantum mechanics [2-6] to quantum communications [7,8] or photonic quantum logic gate [9,10]. Another type of quantum interference is that the interferometric fringe shows the sum of the frequencies of the constituent photons, which is known as NOON state interference [11]. This type of quantum interference has evolved through high-precision phase measurement [11-13] or high-resolution imaging [14]. These two types of quantum interference have been studied independently, but it is possible to explain all the phenomena by calculating the higher-order correlation function.

We recently proposed a connection between HOM or NOON state two-photon interference fringes and the spectral information, and this connection is underpinned by the Wiener–Khinchin formalism [15]. While the conventional Wiener–Khinchin theorem [16,17] is dedicated to the first-order correlation function, we termed the relation with a second-order correlation function for the two-photon quantum interference phenomena an extended Wiener–Khinchin theorem (e-WKT) [15]. This formalism provides a comprehensive and intuitive understanding of two-photon interference; the Fourier transform of the HOM (NOON) fringe gives the difference- (sum-) frequency spectrum of the two constituent photons. Furthermore, the formalism could enable us to accelerate the exploitation of spectroscopic measurements with two-photon interference. Hereafter, we refer to the interferometric measurement based on e-WKT as quantum Fourier transform spectroscopy (QFTS).

In our previous experiment, we demonstrated a proof-of-principle experiment of the e-WKT utilizing photon pairs from spontaneous parametric down-conversion (SPDC) in a nonlinear crystal. The physical process of SPDC is relatively simple and easy to handle; the resultant spectral characteristics from the e-WKT can be understood by the properties of the pump pulse spectrum and the dispersion relation of the nonlinear crystal [18]. On the other hand, a feasibility study with a complex physical system, such as condensed matter, is desirable for the further development of QFTS as a form of practical spectroscopy. To this end, here we present the quantum Fourier transform spectroscopy of a biexciton in a single crystal of the semiconductor CuCl.

**Photon-pair emission via biexciton in CuCl**

CuCl single crystal provides a simple exciton band structure, and thus its exciton properties have been intensively investigated since the early stage of solid-state spectroscopy [19-20]. Furthermore, CuCl is known as the material used for the first entangled-state generation from a semiconductor [21], while many works using semiconductor quantum dots have been reported [22-24]. Unlike entangled-photon generation in semiconductor quantum dots, entangled-state generation in CuCl is obtained via resonant hyper-parametric scattering (RHPS) accompanied by the phase-matching condition in the transition between biexciton and exciton-polariton bands: a biexciton is created by resonant two-photon excitation and then coherently decays into two exciton-polaritons. These polaritons propagate inside the crystal in the direction determined by the phase-matching condition and subsequently convert into photons at the crystal surface. In what follows, we refer to a photon from a higher- (lower-) energy polariton as a HEP (LEP) photon. The presence of the phase-matching condition makes it easy to collect photon pairs at certain emission angles, similar to photon-pair generation with nonlinear crystals [25]. Such well-characterized excitonic properties and photon-pair generation in CuCl can offer a superior environment for the demonstration of QFTS as the first step toward a new solid-state spectroscopy with quantum interferometry.

The experimental setup for the generation and collection of photon pairs is depicted in Fig. 1a. CuCl was pumped by a frequency-doubled mode-locked Ti:sapphire laser with a repetition rate of 76 MHz. The center wavelength of the second harmonic light was tuned at 389 nm with a bandwidth of 0.1 nm for the resonant two-photon excitation of the biexciton. The sample was kept at a temperature of 3.9 K, and the pump beam irradiated CuCl with normal incidence. Then the photon pairs, symmetrically emitted at an angle of 20° to the propagation direction of the pump beam, were collected into polarization-maintaining fibers (PMFs). Figure 1b shows the spectra of the photons scattered into the upper optical path in Fig. 1a with an average pump beam power of 4.0 μW, taken by a CCD camera with a grating spectrometer. The peaks indicated by HEP and LEP are the spectra of photon pairs, and the Rayleigh scattered light of the pump beam appears between the HEP and LEP spectra. The other peaks, labeled $M_T$, $M_L$, and HEP', also originated from the biexciton luminescence but do not form a two-photon state. Thus, in this work we focus only on the spectral peaks of HEP and LEP. A previous report demonstrated the generation of entanglement in the polarization degree of freedom [21], but the photon pairs from the biexciton in CuCl are expected to have entanglement in the frequency degree of freedom simultaneously. Therefore, the two-photon state observed in paths 1 and 2 might be in hyper-entanglement as follows: $(|H\rangle_1|H\rangle_2 + |V\rangle_1|V\rangle_2) \otimes (|\omega_{\text{HEP}}\rangle_1|\omega_{\text{LEP}}\rangle_2 + |\omega_{\text{LEP}}\rangle_1|\omega_{\text{HEP}}\rangle_2)$, where $H$, $V$, $\omega_{\text{HEP}}$, and $\omega_{\text{LEP}}$ represent the states for the horizontal, vertical, and angular frequencies of HEP and LEP photons (indicated by $i = 1,2$), respectively. Such inherent characteristics of the hyper-entangled state via RHPS will be discussed elsewhere. In the present study, since only the frequency entanglement is needed to observe the two-photon quantum interference, we placed polarizing beam splitters (PBSs) before the PMFs.

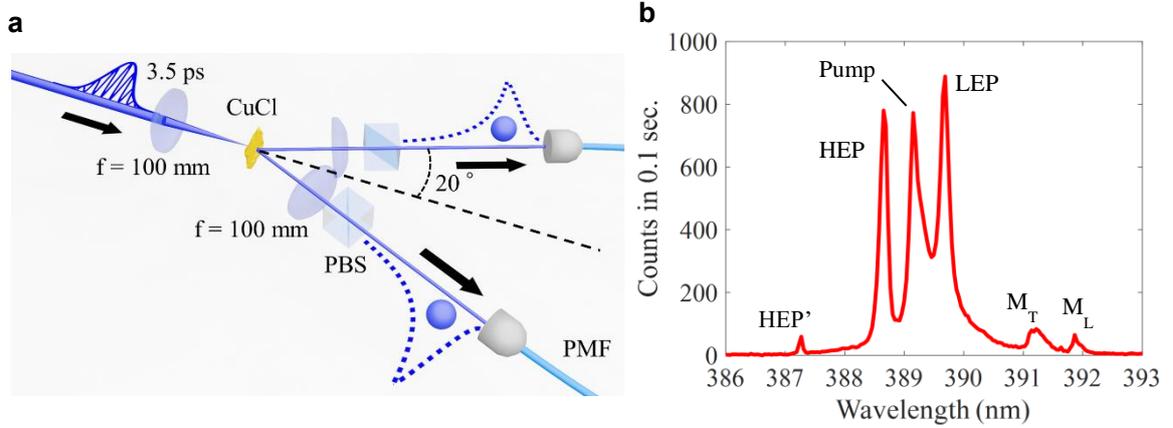

**Fig. 1 | Photon-pair generation in CuCl single crystal.** **a,** Schematic of the generation of photon pairs and their collection into polarization-maintaining fibers (PMFs). Polarizing beam splitters (PBSs) work not only to select a frequency-entangled state but also to reduce the Rayleigh scattered pump beam. **b,** Emission spectrum of the CuCl at 3.9 K with two-photon resonant excitation. The peaks of HEP and LEP are the photon-pair spectra via RHPS. The residual pump beam (Pump) is observed between the HEP and LEP spectra. The two peaks of $M_T$ and $M_L$ in the lower-energy side are the biexciton luminescences leaving the transverse and longitudinal excitons, respectively. HEP' also originates from RHPS but does not form a two-photon state.

**Two-photon quantum interferometer**

For the HOM and NOON state interferometry, we constructed a polarization-mode, not a standard-path-mode, Mach−Zehnder interferometer (MZI), as shown in Fig. 2a. The use of polarization modes provides a versatile interferometer setup just by rotating the half-waveplates (HWPs). Here the quarter-wave plates (QWP1 and 2) inside the interferometer were fixed at the optical axes of 22.5° to the horizontal axis. When all the optical axes of the HWPs were set to 0°, no interferometric effects were observed. We used this configuration to evaluate the production rates of the photon pairs. Setting the angle of HWP2 to 22.5° and keeping the other plate at 0°, we observed the HOM fringes by scanning the arrival time of one of the photons. Besides, all the waveplates are set to 22.5°; the interferometer works as a NOON state interferometer by scanning the relative delay $\Delta\tau$. Here we emphasize that no spectral filter was used in any of the measurements because we intended to demonstrate the extraction of spectral information only by the QFTS. Thus, all the spectral components, as shown in Fig. 2b, contributed to the photon counting measurements.

We oriented the axes of PMFs so that the polarization of the two photons was orthogonal to each other at the input side of the interferometer. After the spatial modes of the two photons were combined by a polarizing beam splitter (PBS1), photon pairs were fed into the polarization-mode MZI. Outputted photon pairs from the interferometer were coupled into multimode fibers (MMFs) after the spatial mode was divided by a polarizing beam splitter (PBS3). We then performed coincidence measurements with two single-photon avalanche diodes (SPADs) and a time-interval analyzer.

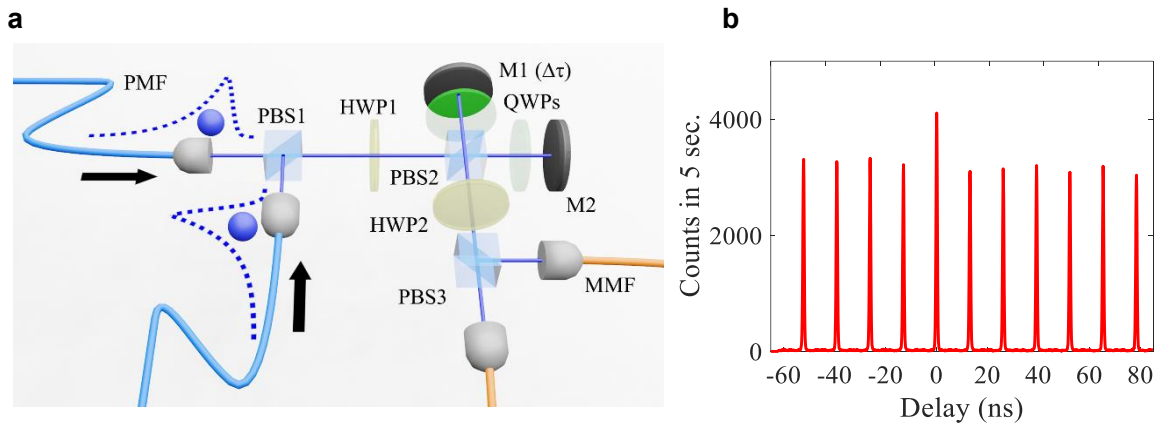

**Fig. 2 | Two-photon quantum interferometer. a,** Schematic of the polarization-mode Mach−Zehnder interferometer. The time delay $\Delta\tau$ for the HOM and the NOON state interference were controlled by scanning the position of the mirror M1. **b,** Time correlation histogram of photon pairs through the interferometer when setting the half-wave plates (HWP1 and HWP2) to 0°. The central peak at the delay time of 0 contains photon-pair signals.

Figure 2b shows the histogram of the time-interval analysis of the photon counting signals from the two SPADs. We saw slightly larger peak counts at the delay time of 0 ns and estimated the accidental coincidence counts in the peak at the 0-delay position from the counting values in the side peaks. Then, subtracting the accidental counts as background, we recorded the coincidence counts caused by the photon pairs.

The spectroscopic capability of the coincidence counting measurement is worth discussing. The coincidence measurement works to extract only the photon-pair spectral components. In our case, we recorded the coincidence counting rate of ~1600 counts/sec against the single-photon counting rate of ~650000 counts/sec, meaning a signal-to-noise ratio of 0.25%. Even under such conditions, we could select the spectra of the HEP and LEP apart from the other biexciton luminescences and the Rayleigh scattered light of the pump beam, without any spectral filters, by only coincidence measurements. Therefore, we can consider the coincidence counting measurement as "photon-number-resolving spectroscopy". Photon-number-resolving measurement provides a powerful tool not only for quantum information technologies but also for spectroscopic quantum measurements.

**Fourier transform spectroscopy by quantum interferometry**

We start by discussing the HOM interference experiment in advance of the two-photon NOON state interference because the NOON state interference is performed as the extension of the HOM interference. Figure 3a presents the resultant HOM interference fringe pattern with a visibility of 61% in our experiment. Here we emphasize that this is the first experimental observation of the HOM interference of photon pairs generated from biexciton luminescence, even though HOM experiments with single photons from semiconductor quantum dots were conducted to test the indistinguishability [26]. The visibility value is somewhat lower than that by photon pairs from SPDC but high enough to perform Fourier transform spectroscopy. According to the e-WKT, we applied the fast Fourier transform (FFT) to our data and then obtained the difference-frequency spectrum of photon pairs from the biexciton, as shown in Fig. 3b. The peak position of the spectrum, 2.03 THz, is associated with the oscillation frequency of the HOM that originated from the frequency difference between the HEP and LEP photons. In practice, the spectral peak separation between the HEP and LEP in Fig. 1a is 2.0 THz and agrees nicely with the Fourier transformed spectrum. These peak positions can be estimated by the phase-matching condition for RHPS [27], implying that the dispersion relation of the exciton-polariton would be dominant for the characteristics of the difference-frequency spectrum. Besides, the coherence time in the HOM is relevant to the spectral width of the difference-frequency spectrum. This spectral width value of 0.43 THz is mostly twice that by the HEP or LEP in Fig. 1b because the FFT spectrum is provided as a function of the difference frequency between the HEP and LEP photons.

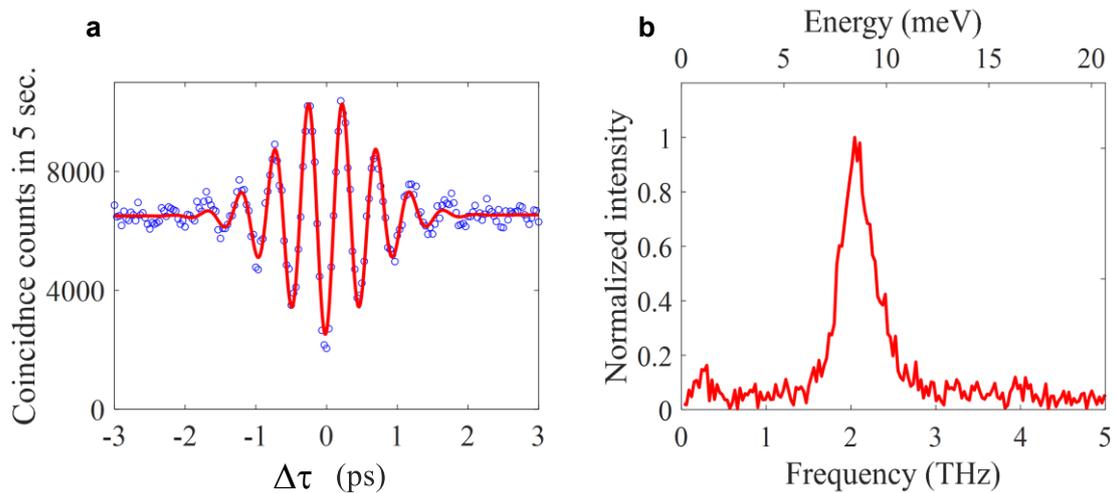

**Fig. 3 | Fourier transform spectroscopy with HOM interference. a,** Two-photon coincidence counts as a function of the delay of $\Delta\tau$ when HWP1 and HWP2 are set to 0° and 22.5° in Fig. 2a, respectively. This setup provides the HOM interference pattern. Open circles are the measured data, and the solid curve is obtained from the fitting. **b,** The difference-frequency spectrum of photon pairs by making the fast Fourier transform of the HOM patterns in **a**.

Following the HOM experiment, we carried out the NOON state interference experiment. Figure 4a represents the fringe pattern around the 0-delay position of the interferometer. Fitting the pattern with a sinusoidal function, we estimated an oscillation period of 173±6 nm with a visibility of 82%. From the viewpoint of the e-WKT, this period means the sum of the frequencies of the constituent photons. In contrast, the corresponding two-photon energy, 7.2±0.3 eV, derived from the sum frequency is close to the biexciton energy of 6.3722 eV in CuCl [28], implying that the fringe pattern by the NOON state interferometry purely reflects the properties of the biexciton. For further spectral analysis, we carried out a long-range interferometric measurement to observe the envelope shape of the fringe pattern. In this measurement, we took coincidence counts for 5 seconds by scanning every 5 μm. Ideally, the scanning step should be a much shorter value than the oscillation period, but we had technical difficulties in retaining the stability of the NOON state interferometer during the long measurement time. Thus, we observed the envelope shape separately from the oscillation period, assuming that only the single sum-frequency oscillation would contribute to the interference pattern. Figure 4b shows the observed pattern with coarse scanning for the long-range measurement. In SPDC experiments, it is well known that the coherence time that appeared in the NOON state interference is relevant to that of the pump pulse. However, the coherence time of the interference in Fig. 4b is approximately 17 ps, which is much longer than the pump pulse duration of 3.4 ps. In a previous experiment on the biexciton in CuCl, the dephasing time of the biexciton measured at 4 K was reported as 16 ps [28] by nonlinear spectroscopy of four-wave mixing. Since the dephasing times of the biexciton in CuCl varied from 16 to 50 [29,30] ps from sample to sample, in the previous experiments, the coherence time observed in the NOON state interferometry could be considered to reflect the dephasing time of the biexciton.

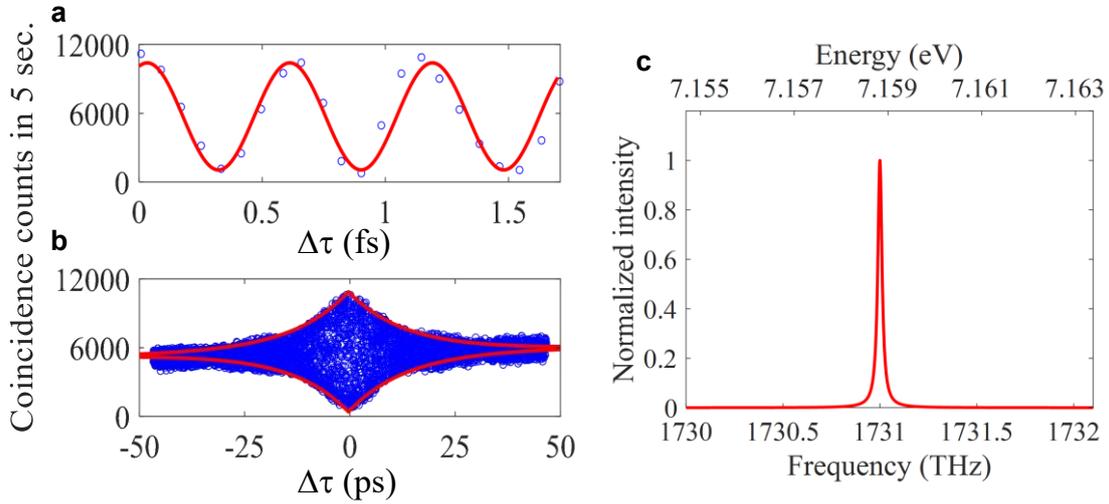

**Fig. 4 | Fourier transform spectroscopy with NOON state interference.** Two-photon coincidence counts as a function of the delay of $\Delta\tau$ when all the HWPs are set to 22.5° in Fig. 2a, providing the NOON state interference pattern around the 0-delay position (**a**) and in the long-range scanning (**b**). The solid curve in **a** is the fitting for the interferometric oscillation and that in **b** for the envelop shape. **c,** the sum-frequency spectrum of photon pairs obtained from the Fourier transform of the reproduced interference pattern from the fitting curves.

For the Fourier transform spectroscopy, we assume the following function in order to reproduce the interference pattern: $I(\tau) = e^{-\gamma|\tau|}(1 + V\cos\omega\tau)$, where $I$, $\gamma$, $V$, and $\omega$ are the normalized counting rate, decay rate, visibility, and oscillation frequency, respectively. We estimated the oscillation frequency $\omega$ and the visibility $V$ from the fine scanning data around the 0-delay position in Fig. 4a. On the other hand, the decay rate $\gamma$ was determined by fitting the envelope shape in Fig. 4b. The exponentially decaying function was used in the previous time-resolved measurements of biexciton luminescence [30] and well fitted our observed data indeed. Then, applying the Fourier transform to the reproduced NOON state interference pattern, we estimated the sum-frequency spectrum, as shown in Fig. 4c. As discussed above, the center frequency and spectral shape would be purely reflected in the biexciton properties. Additionally, the spectral width of 0.11 meV is much narrower than that of the HEP or LEP spectrum, indicating that it is hard to observe the sum-frequency spectrum by classical spectroscopic measurements. Therefore, we could say that Fourier transform spectroscopy with NOON state interferometry allows us to measure the biexciton spectrum reflecting the intraband relaxation, thereby distinguishing it from the spectrum determined by the interband transition between the biexciton and the exciton-polariton.

**Discussion**

Conventionally, the single-photon emission caused by the transition from a biexciton state to an intermediate state was previously measured to elucidate biexciton characteristics [29-32]; in our case, the exciton-polariton describes the intermediate state. The characteristics of a single photon, however, are affected by not only the biexciton state but also the intermediate state. Thus, single-photon spectral measurement cannot observe a pure spectrum reflecting the intraband relaxation of a biexciton. On the other hand, as demonstrated in this paper, two-photon spectral measurement allows the measurement of the biexciton spectrum separated from the influence on the intermediate state. This unique feature of QFTS, based on the measurement axes of the sum or difference frequencies, has great potential for two-photon nonlinear spectroscopy at the single-photon level.

It should be noted that two-photon quantum interferometry requires a symmetric state for interchanging the constituent photons in any degree of freedom. Since photon pairs from the biexciton in CuCl naturally satisfy the symmetry condition, we can successfully demonstrate biexciton spectroscopy with QFTS. Nevertheless, the requirement of symmetry might limit the applicability of QFTS to candidate materials. However, even if photon pairs do not satisfy the symmetry condition, we might overcome the limitation by creating a superposition state of the two-photon state. Indeed, the symmetry condition can be realized by placing a material inside an interferometer in some quantum optical experiments with nonlinear crystals [33]. This would mitigate the exchange symmetry condition for QFTS. Our next challenge will be to carry out QFTS for two-photon emission without the symmetry condition in order to establish its versatility.

Additionally, the presence of the phase-matching condition in the RHPS process of CuCl makes it easy to collect photon-pair emissions into optical fibers. However, two-photon emission without any phase-matching condition would often be observed in solid-state spectroscopy, and then the two-photons are randomly emitted in any directions due to the absence of momentum correlation between the photons. Thus, the development of a means to collect two photons in a wide solid angle would also be crucial for applying QFTS to investigate various condensed matters.

**Conclusions**

We demonstrated Fourier transform spectroscopy with quantum interferometry for photon pairs produced via the CuCl biexciton. First, we discussed the applicability of two-photon counting measurement to spectroscopy. Consequently, we showed that the photon-number-resolving measurement could work well as it extracts only the spectral component of the photon pair in a high signal-to-noise ratio and has great potential for the spectroscopy of condensed matter systems. Secondly, we demonstrated that the two-photon quantum interference provides two types of interferometric waveforms: one gives the fringe with the difference frequency and the other that with the sum frequency of the constituent photons. The Fourier-transformed difference-frequency spectrum can be characterized by the phase-matching condition associated with the dispersion properties of the exciton-polariton. In contrast, the QFTS with the two-photon sum frequency allows us to unveil the spectrum that purely reflects the intraband relaxation of the biexciton. Generally, two-photon emission exhibits an intricate spectrum because of the contributions of multiple quantum states. However, QFTS allows us to observe the two-photon spectrum as a function of the sum or difference frequency, making it simple to study a composite quantum system, such as a biexciton, as decomposed into each associated state. These remarkable features of QFTS constitute a new tool to investigate nonlinear light−matter interactions with two-photon emission.

**Methods**

**Extended Wiener−Khinchin theorem.** The two-photon detection probabilities for the NOON state interference $P_2^+$ and for the HOM interference $P_2^-$ are given by the following form;

$$P_2^\pm(\tau) = \frac{1}{2}\left\{1 \pm \mathrm{Re}\left[\int_{-\infty}^{\infty}\int_{-\infty}^{\infty} d\omega_1 d\omega_2 \, |f_2(\omega_1,\omega_2)|^2 e^{-i(\omega_1\pm\omega_2)\tau}\right]\right\}, \quad (1)$$

where $f_2(\omega_1,\omega_2)$ is the two-photon spectral probability amplitude for the photons with angular frequencies $\omega_1$ and $\omega_2$. The symmetry condition, $f_2(\omega_1,\omega_2) = f_2(\omega_2,\omega_1)$, is needed for this expression, but we can expect that the photon pairs from CuCl would be satisfied. Here we introduce the sum-frequency spectrum $F_2^+$ and the difference-frequency spectrum $F_2^-$ as $F_2^\pm(\omega_\pm) \equiv \int_{-\infty}^{\infty} d\omega_\mp |f_2(\omega_+,\omega_-)|^2$, where $\omega_\pm = \omega_1 \pm \omega_2$. Then, the two-photon detection probabilities are rewritten as

$$P_2^\pm(\tau) = \frac{1}{2}\left\{1 \pm \mathrm{Re}\left[\int_{-\infty}^{\infty} d\omega_\pm F_2^\pm(\omega_\pm) e^{-i\omega_\pm \tau}\right]\right\}. \quad (2)$$

This equation describes that the Fourier transform of the sum- or difference-frequency spectrum provides the two-photon quantum interference pattern. Moreover, we can define the second-order correlation function as the Fourier transform of the $F_2^+$;

$$G_2^\pm(\tau) \equiv \int_{-\infty}^{\infty} d\omega_\pm F_2^\pm(\omega_\pm) e^{-i\omega_\pm \tau}. \quad (3)$$

For the inverse Fourier transform, we obtain

$$F_2^\pm(\omega_\pm) = \frac{1}{2\pi}\int_{-\infty}^{\infty} d\tau \, G_2^\pm(\tau) e^{i\omega_\pm \tau}. \quad (4)$$

These Fourier transform relations between the $F_2^+$ and the $G_2^\pm$ are the extended Wiener−Khinchin theorem, which underpins the quantum Fourier transform spectroscopy presented in this paper.

**Sample fabrication.** The samples, CuCl crystals, were grown by a vapor phase transport method. CuCl powders (Sigma Aldrich; purity of 99.9995%) were put on a quartz boat, which was then inserted in a quartz tube. The quartz tube was then filled with Ar gas of 0.3 atm, then sealed. Using an electric furnace, the CuCl powders were heated at 430 ℃ for approximately 14 days, during which CuCl single crystals grew as platelets on the tube's internal wall. The thickness of the CuCl platelets was approximately several micrometers.

**Acknowledgements**

This work was supported by MEXT Quantum Leap Flagship Program (MEXT Q-LEAP) Grant Number JPMX S0118069242.

Contributions

R.S. conceived the concept behind the research and supervised this project. K.M. fabricated the samples. H.S experimented with all the optical measurements and analyzed the data. All authors contributed to writing the manuscript.

Correspondence

Correspondence to Ryosuke Shimizu (r-simizu@uec.ac.jp).

Competing Interests

The authors declare no competing interests.